# Advancing atomic electron tomography with neural networks

Juhyeok Lee[1,2,†] and Yongsoo Yang[3,4*]

[1] *Energy Geosciences Division, Lawrence Berkeley National Laboratory, Berkeley, CA 94720, USA*
[2] *National Center for Electron Microscopy, Lawrence Berkeley National Laboratory, Berkeley, CA 94720, USA*
[3] *Department of Physics, Korea Advanced Institute of Science and Technology (KAIST), Daejeon 34141, Republic of Korea*
[4] *Graduate School of Semiconductor Technology, School of Electrical Engineering, Korea Advanced Institute of Science and Technology (KAIST), Daejeon 34141, Republic of Korea*

Corresponding authors, email: [†] jhlee0667@lbl.gov, [*] yongsoo.yang@kaist.ac.kr

**Abstract**
Accurate determination of three-dimensional (3D) atomic structures is crucial for understanding and controlling the properties of nanomaterials. Atomic electron tomography (AET) offers non-destructive atomic imaging with picometer-level precision, enabling the resolution of defects, interfaces, and strain fields in 3D, as well as the observation of dynamic structural evolution. However, reconstruction artifacts arising from geometric limitations and electron dose constraints can hinder reliable atomic structure determination. Recent progress has integrated deep learning, especially convolutional neural networks, into AET workflows to improve reconstruction fidelity. This review highlights recent advances in neural network-assisted AET, emphasizing its role in overcoming persistent challenges in 3D atomic imaging. By significantly enhancing the accuracy of both surface and bulk structural characterization, these methods are advancing the frontiers of nanoscience and enabling new opportunities in materials research and technology.

**Introduction**
In recent years, the demand for innovative nanomaterials and nanostructures has surged across diverse fields, including catalysis[1–9], electronics[10–15], energy storage[16–22], quantum technologies[23–28], structural materials[29–34], biosensing[27,35–38], and targeted drug delivery[39–42]. Designing materials with tailored functionalities requires precise control over the arrangement of atoms—the fundamental building blocks of matter[43–46]. While exceptions exist, such as atomic chains and purely two-dimensional (2D) materials like monolayer graphene, most materials, even those classified as 0-dimensional, 1-dimensional, or 2D, possess inherently three-dimensional (3D) atomic arrangements. To fully manipulate these structures, it is essential to not only measure atomic positions with high precision but also understand how they evolve under external influences such as temperature, electromagnetic fields, and pressure.

For decades, transmission electron microscopy (TEM) has been a primary tool for atomic-scale structural analysis; more recently, the advent of aberration correctors has significantly enhanced achievable resolution, making atomic resolution imaging routinely attainable[47–54]. However, TEM images provide only 2D projections of 3D structures, limiting their ability to capture atomic arrangements in all three dimensions[55–60]. Scanning probe techniques can resolve surface features at the atomic scale but lack access to subsurface structures[61–66]. Crystallographic methods using X-rays, electrons, or neutrons enable high-resolution 3D structural determination[67–73] but are largely restricted to periodic crystals, making them unsuitable for studying non-crystalline features such as grain boundaries, dislocations, interfaces, and point defects[69,70,73,74]. Other techniques, like coherent diffractive imaging, have shown promise for 2D and 3D analysis but have not yet achieved true atomic resolution[75–81]. While atom probe tomography can provide 3D atomic information, it is a destructive technique and is challenging to apply for dynamic studies[82–86].

Electron tomography has emerged as a powerful tool for non-destructive 3D atomic structural analysis[87–89]. In 2012, it achieved near-atomic resolution (2.4 Å) without relying on crystalline symmetry[90]. In 2015, atomic electron tomography (AET) was demonstrated[91–93], achieving an atomic-level precision of 19 picometers by combining aberration-corrected TEM with advanced iterative reconstruction algorithms[93,94]. This milestone enabled the direct determination of 3D atomic positions of individual atoms in nanomaterials, shifting AET from qualitative visualization to precise, quantitative materials characterization (Fig. 1a-c).

Since then, AET has been widely used to investigate complex non-crystalline atomic structures such as grain boundaries[95,96], dislocations[97,98], stacking faults[60,90,98], point defects[93,95], local distortions[99–101], heterointerfaces[100,102–105], amorphous structures[98,106,107], and strain tensors[60,91,93,95,103,108–110] in unprecedented 3D

detail (Fig. 1d-i). Leveraging its non-destructive nature, AET has also enabled 4D (3D space + time) atomic-resolution imaging, capturing dynamic structural changes over time, such as nucleation and growth processes[111–113]. Furthermore, experimentally determined atomic coordinates have been integrated into ab initio calculations, allowing direct correlations between atomic-scale structures and physical, chemical, and electronic properties of various materials[43,90,93,95,99,108,109,114–116].

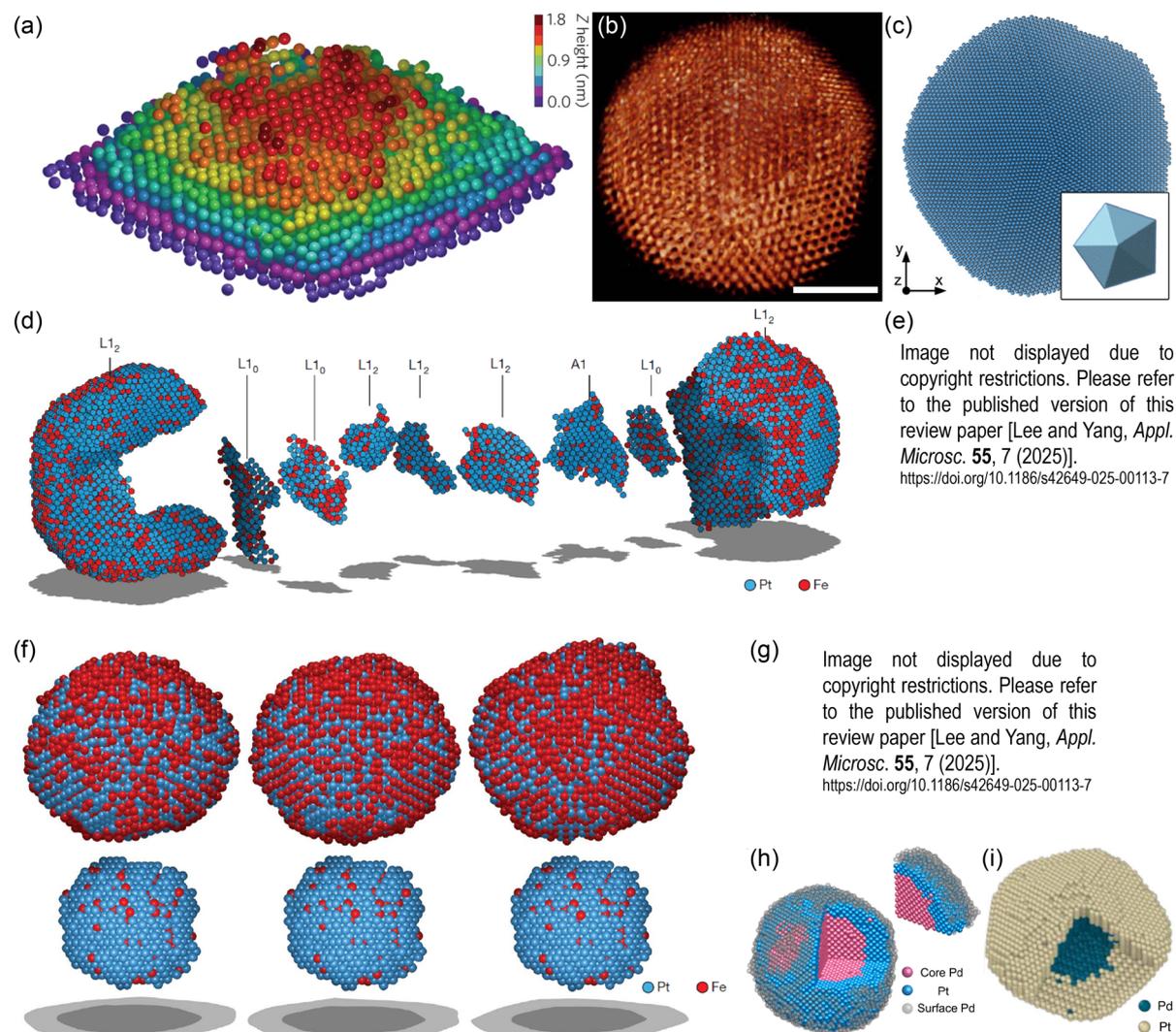

**Figure 1 | Representative 3D atomic structural analysis based on electron tomography.** (a) 3D positions of individual atoms in a tungsten needle sample. The 3D atomic structure consists of nine atomic layers along the [011] direction, labelled with dark red, red, orange, yellow, green, cyan, blue, magenta and purple from layers 1–9, respectively. Adapted from Reference [93], © 2015 Springer Nature. (b) A three-dimensional reconstruction of a Ag-Au nanocluster, showing atomic structure and composition of the cluster. Scale bar, 2 nm. Adapted from Reference [92], © 2015 Macmillan Publishers Limited. (c) 3D visualization of a reconstructed Au nanodecahedron containing more than 90 000 atoms. Adapted from Reference [91], this is an unofficial adaptation of an article that appeared in an ACS publication. ACS has not endorsed the content of this adaptation or the context of its use. © 2015 American Chemical Society. (d) Experimentally determined complex grain structure of an FePt nanoparticle via atomic electron tomography. Adapted from Reference [95], © 2017 Springer Nature. (e) Experimental 3D atomic model of an amorphous nanoparticle composed of eight chemical elements (Reference [106]). (f) 3D atomic models of an FePt nanoparticle after 9 minutes (left), 16 minutes (middle), and 26 minutes (right) of accumulated annealing. The top row shows the entire nanoparticle, while the bottom row highlights the Pt-rich core at each stage. Adapted from Reference [111], © 2019 Springer Nature. (g) 3D density maps and atomic positions of a single-crystalline Pt nanocrystal along the [111] zone axis. Scale bar, 1 nm (Reference [110]). (h-i) Experimentally determined 3D atomic structures of Pd@Pt core-shell nanoparticles, revealing (h) strain correlation between the surface and interface and (i) chemical diffusion at the interface. Adapted from References [103,104].

Despite its success, AET faces a major challenge stemming from geometric constraints in electron tomography experiments, where the specimen holder or grid obstructs the electron beam beyond certain tilt angles, preventing the full acquisition of angular data[58,117]. This limitation, known as the "missing wedge" problem, introduces elongation along beam direction and Fourier ringing artifacts in reconstructed tomograms, leading to distortions that particularly affect surface atomic structures[117–119]. As a result, while AET has been widely applied to study internal atomic arrangements, achieving precise 3D surface atomic structure determination remains a persistent challenge.

This limitation is especially critical because surface atomic configurations govern key material properties and applications. For example, catalytic activity is almost entirely dictated by the arrangement of surface atoms, rather than the internal atomic structure[120]. Likewise, surface atomic structures influence adhesion[121], corrosion resistance[122], electronic transport[123,124], and interfacial phenomena[125,126] in a wide range of materials. Accurately determining the 3D atomic structure of surfaces is thus essential for both fundamental scientific insights and practical applications.

To address the missing wedge problem, various approaches have been explored, including deep learning-based neural networks. Convolutional neural networks (CNNs), in particular, have gained significant attraction in electron microscopy for tasks such as missing data reconstruction and super-resolution imaging[127–131]. Recent advancements have demonstrated the successful integration of AET with neural networks guided by the atomicity principle (which assumes that the sample consists solely of discrete atomic potentials), substantially improving the reliability of surface 3D atomic structural characterization[108,132–135].

This review highlights recent advancements in AET, with a special focus on neural network-based methodologies. We discuss fundamental concepts behind neural network-assisted atomic-resolution tomography, explore experimental and computational success cases, and introduce novel approaches such as image inpainting for improving tomography tilt series before reconstruction. By leveraging machine learning, AET is poised to overcome long-standing limitations. This approach offers unprecedented precision in 3D atomic structure determination and paves the way for breakthroughs in materials science and nanotechnology.

**Deep learning for missing wedge recovery in AET**

Recent advances have addressed the long-standing issue of missing wedge artifacts in AET by integrating machine learning to recover unmeasured information and enhance reconstruction quality[60,108,112,131–135]. These approaches typically train neural networks to inpaint missing tilt projections or remove artifacts in reconstructed tomograms, thereby surpassing the limitations of conventional reconstruction algorithms. A notable example is the two-step deep learning pipeline developed by Ding et al., which first infills missing wedge data in the sinogram domain (a volume of collected tilt series) and subsequently refines the 3D reconstruction using a U-Net-based architecture[132]. The first step employs a generative adversarial network (GAN) to predict the data corresponding to missing tilt angles in the sinogram (Fig. 2a), while the second applies an encoder-decoder network with skip connections to suppress residual artifacts in the tomogram. This joint model produced reconstructions of significantly higher fidelity than traditional weighted back-projection[136] or simultaneous algebraic reconstruction technique[137] algorithms, even when over 80% of the tilt range was missing. The method was later extended to the information recovery and de-artifact model (IRDM), which achieved sub-angstrom resolution (0.7 Å) in the electron tomography of nanoporous gold[133]. These results convincingly demonstrate that deep learning can effectively mitigate the missing wedge problem in AET.

A further advancement was introduced in 2021, in which a 3D neural network was inserted as a post-reconstruction augmentation step in the AET reconstruction pipeline[108]. Lee et al. implemented a 3D U-Net architecture[108,138] to enhance preliminary reconstructions obtained from iterative algorithms such as GENFIRE[139] (Fig. 2b). The initial "raw" tomograms often suffer from blurring and elongation due to the missing wedge. These tomograms were processed through the trained U-Net, resulting in refined volumes. In the output, atomic sites appear as well-isolated peaks with the expected Gaussian intensity profiles. The tomogram augmentation approach also introduced a unique and powerful prior, known as the atomicity constraint, which is specifically tailored for atomic electron tomography. This prior is based on the assumption that the sample consists of discrete atomic potentials. As a result, the network learned to transform blurred density distributions into well-resolved atomic peaks, even when applied to structures that were entirely different from those used during training. This approach led to a substantial increase in reconstruction accuracy, achieving an atomic coordinate precision of 15.1 pm.

More recently, Yu et al. introduced further improvements by developing an ensemble cross U-Net transformer model (EC-UNETR)[134] (Fig. 2c). The EC-UNETR design incorporates hierarchical subnetworks that process features along orthogonal axes and includes a skip fusing unit to improve network stability and flexibility. Transformer-based attention mechanisms are embedded within both the encoder and decoder stages, enabling the model to capture long-range spatial dependencies. This architecture improved the recovery of fine structural details, reduced reconstruction artifacts, and yielded a 5.5% decrease in root-mean-square error compared to earlier models.

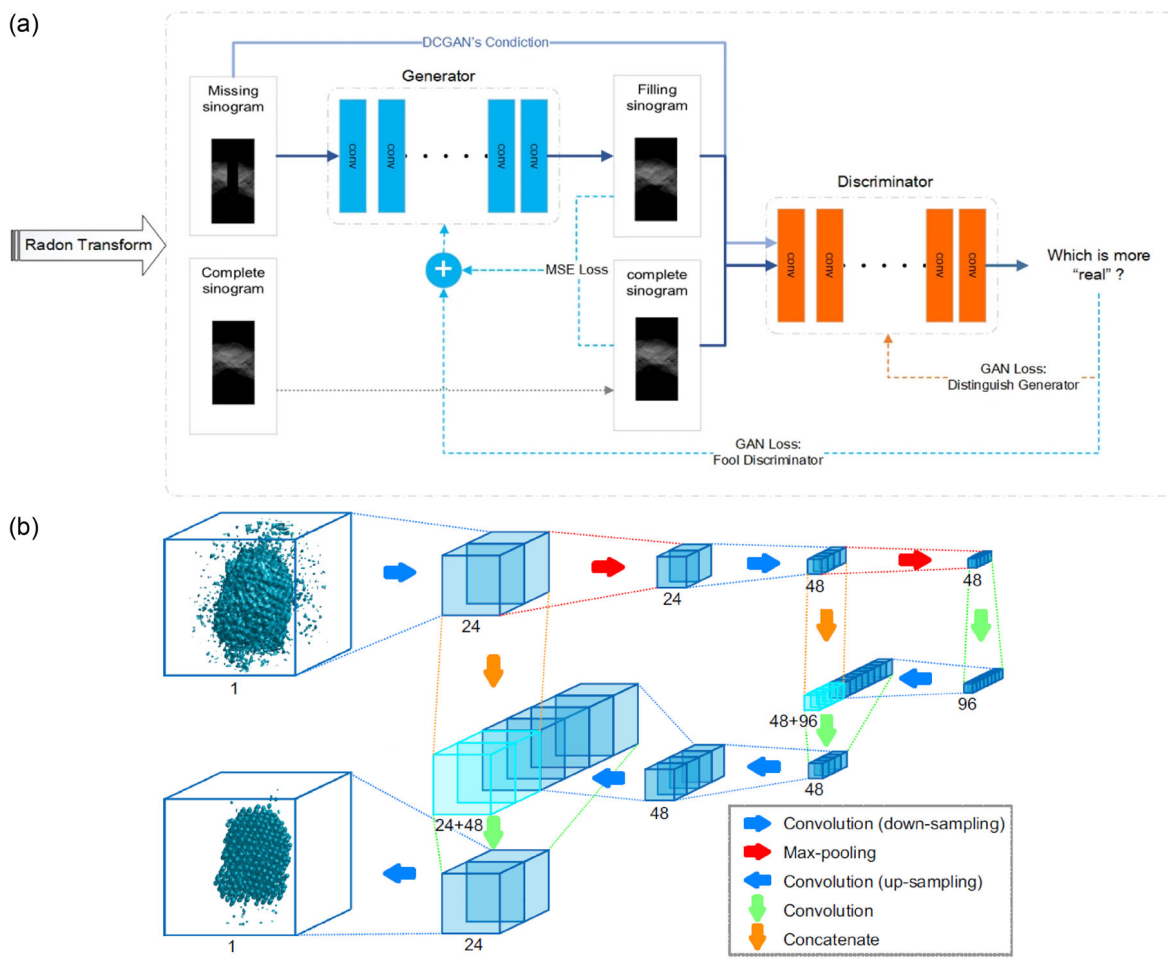

**Figure 2 | Neural network architectures for enhancing AET.** (a) A residual-in-residual dense block-based GAN that fills missing regions in the sinogram domain as the first step of a two-stage process, followed by a U-Net-based network for artifact reduction in the reconstructed volume. Adapted from Reference [132]. (b) Architecture of the deep learning augmentation applied to 3D tomograms. The model follows a 3D U-Net structure, where each box represents a feature map. The number of channels is indicated below each feature map. Adapted from Reference [108]. (c) Overview of the skip fusing unit within the EC-UNETR framework. The outputs from the preceding subnetworks within the identical stage and the upsampled outputs from the lower stage are concatenated and normalized to calculate the fusing weights (Reference [134]).

**Precise determination of Pt nanoparticle surfaces and interface structures**
The application of 3D U-Net-based augmentation has enabled reliable determination of three-dimensional surface atomic structures of platinum nanoparticles at single-atom resolution. In the first demonstration, nearly 1,500 atoms (including low-coordination surface atoms) were precisely located and identified within a single Pt nanoparticle[108] (Fig. 3a-h). The neural network augmentation significantly improved both atom detectability and positional accuracy (Fig. 3a-f): the fraction of correctly identified atoms increased from approximately 96.5% to 98.8%, and the root-mean-square deviation in atomic coordinates decreased from 26.1 pm to 15.1 pm. To validate the reliability of the determined atomic model, R-factor was computed by comparing the experimental tilt series to calculated projections derived from the reconstructed atomic structures. The R-factor improved from 19.2% to 17.4%. The lower R-factor for the deep learning-augmented tomogram indicates greater consistency with the experimental data, affirming that the reconstruction is more faithful to the true structure. Importantly, many surface atoms that were unidentifiable in the uncorrected tomogram due to signal smearing or intensity loss caused by the missing wedge became clearly resolvable after deep learning enhancement. These results allowed for detailed analysis of surface strain fields, revealing facet-dependent strain states: the {100} and {111} surfaces contributed unequally to the overall anisotropic strain distribution, with localized compressive strain identified at the particle-support interface (Fig. 3g-h).

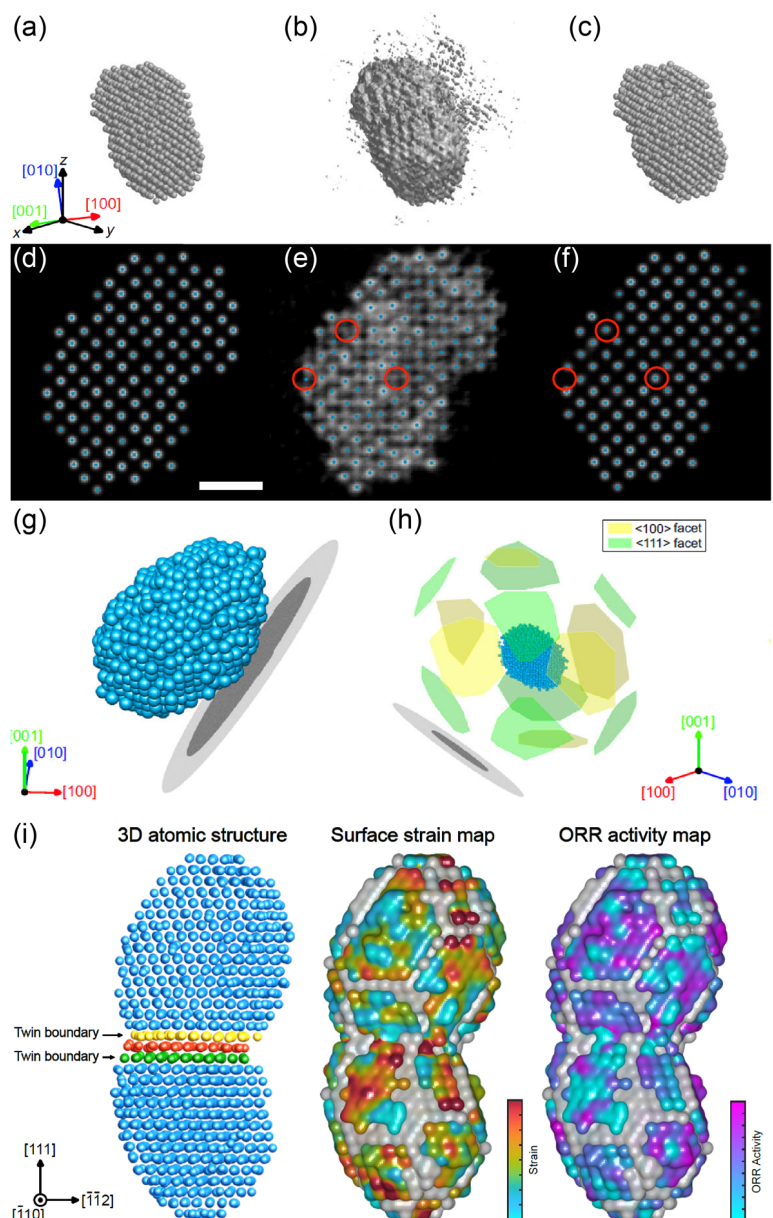

**Figure 3 | Determination of Pt nanoparticle surface and interface structures via deep learning augmentation.** (a-c) 3D iso-surfaces plotted with 10% iso-surface values (10% of the highest intensity), representing ground truth (a), linear tomogram before (b) and after the augmentation (c) from simulation of AET process for a Pt nanoparticle. Note that the z direction is the missing wedge direction. (d-f) 2-Å-thick slices perpendicular to [001] direction, obtained from the 3D tomograms near the center region. Ground truth (d), linear tomogram before (e) and after the augmentation (f). The grayscale background represents the reconstructed intensity, and blue dots represent the positions of traced atoms. Red circles denote misidentified atoms before the augmentation, which become correctly traced after the augmentation. Scale bar, 1 nm. (g) Experimentally determined 3D atomic structures of a Pt nanoparticle obtained via AET with neural network-based augmentation. The SiN substrate appears as black and gray disks. (h) Identified surface facets of the Pt nanoparticle, showing both <100> and <111> facets. Adapted from Reference [108]. (i) 3D structure of a Pt nanodumbbell revealing a twin boundary at the interface. Surface strain, measured at atomic resolution, was directly correlated with catalytic activity through DFT calculations. Adapted from Reference [60] with permission, © 2022 American Chemical Society.

Beyond isolated nanoparticles, neural network-assisted AET also enabled the direct visualization of the interface between coalescing Pt particles. More specifically, a dumbbell-shaped Pt nanoparticle formed by the coalescence of two clusters was reconstructed in full 3D, revealing a distinct double twin boundary at the junction and significant atomic disorder throughout the structure[60] (Fig. 3i). Using the atomic coordinates, a complete three-dimensional strain tensor was derived, uncovering strong tensile strain localized at the protruded {100} region. This experimental strain map was then used as direct input for density functional theory calculations, linking the observed strain to enhanced oxygen reduction reaction activity on the strained facet. These studies underscore the power of neural network-assisted AET in uncovering subtle structural features such as twin boundaries, strain anisotropy, and coalescence-induced disorder, while also establishing quantitative links between structure and catalytic functionality.

It is important to note, however, that the reported surface atomic structures should not be interpreted as perfectly accurate representations of static configurations. At room temperature, surface atoms can undergo spontaneous diffusion at $10^{-4}$–$10^{-7}$ s timescale[140] even without electron beam exposure, and the high electron dose (typically on the order of $10^5$ e Å$^{-2}$) used during AET acquisition may further perturb the surface. These limitations may introduce artifacts into the reconstructed structures. Nonetheless, the obtained surface configurations represent the most detailed experimental insight currently available, providing unparalleled information on the structural behavior of nanomaterial surfaces at atomic resolution.

**Improving AET via image inpainting**

In a recent development, Iwai et al. introduced a CNN-based image inpainting strategy to address challenges in imaging supported metal nanoparticles via AET[135]. Their approach focused on isolating the nanoparticle signal from the complex background contributions of the support material in electron tomography tilt series. The inpainting model effectively predicted and removed the support signal, enabling clearer visualization of the nanoparticle itself (Fig. 4a-c). Using the inpainted dataset, the team reconstructed an 11 nm palladium nanoparticle in 3D, revealing a deformed cuboctahedron structure with high-index facets (Fig. 4d-f). These morphological characteristics are associated with increased catalytic activity, particularly in methane combustion. Furthermore, atom-level mapping of local Pd-Pd bond distances and their variations enabled visualization of interfacial strain and atomic disorder at the Pd/Al$_2$O$_3$ boundary. This study demonstrates the utility of deep learning in preprocessing tilt series data to isolate relevant structures from noisy or overlapping signals, thereby expanding the applicability of AET to supported systems. The ability to accurately reconstruct and analyze nanoparticles in realistic environments opens new pathways for rational catalyst design and performance optimization.

Image not displayed due to copyright restrictions. Please refer to the published version of this review paper [Lee and Yang, *Appl. Microsc.* **55**, 7 (2025)].
https://doi.org/10.1186/s42649-025-00113-7

**Figure 4 | Enhancing AET through image inpainting** (Reference [135]).

## Summary and Outlook

Recent advances in deep learning have significantly expanded the capabilities of AET, enabling reliable three-dimensional imaging of materials with single-atom precision, even under challenging conditions such as limited tilt ranges, strong background interference, and complex surface structures. By integrating convolutional neural networks into the AET reconstruction pipeline, researchers have developed powerful tools for recovering missing information, suppressing reconstruction artifacts, and refining atomic-resolution data.

Tomogram augmentation with neural networks, including U-Net and transformer-based models, has substantially improved reconstruction accuracy and atomic site resolution. By incorporating physical priors such as atomicity, these methods generalize across diverse structural types and have enabled precise mapping of facet, strain, and distortions in nanoparticles. When combined with theoretical modeling, they support quantitative analysis of catalytic activity at the atomic scale. In parallel, CNN-based inpainting techniques have proven effective in isolating nanoparticle signals from complex backgrounds, facilitating 3D analysis of supported catalysts. Collectively, these advances are transforming AET into a versatile, data-driven platform for atomic-scale materials characterization.

Looking ahead, further integration of deep learning with advanced AET methodologies is expected to unlock new frontiers. Conventional AET techniques based on annular dark-field measurements remain limited to relatively heavy elements due to their stronger electron scattering signals. However, many technologically relevant nanomaterials contain low-Z elements (such as oxygen, carbon, and nitrogen) alongside heavier atoms. In particular, for oxide systems, understanding their emergent properties—especially those arising at the nanoscale or at heterointerfaces—requires accurate 3D imaging of light elements such as oxygen[101,141–143].

In this context, four-dimensional scanning transmission electron microscopy (4D-STEM)[144,145], enabled by high-speed pixelated detectors, offers new opportunities by fully exploiting the information carried by scattered electrons in the diffraction plane. Recent experimental and computational developments suggest that 4D-STEM-based ptychography[146–152] and tomography[153–159] can resolve atomic positions of both light and heavy elements with high accuracy. Furthermore, newly developed 4D-STEM multislice tomography methods enable out-of-focus imaging, overcoming depth-of-focus limitations and allowing analysis of larger sample volumes. Similarly, multislice tomography using multiple tilted high-resolution TEM focal series offers comparable benefits. It achieves high-accuracy atomic resolution for both light and heavy elements while mitigating depth-of-focus constraints, making it especially suitable for beam-sensitive materials, such as biological samples[153,160,161]. When combined with neural network-based reconstruction, these developments promise full 3D atomic mapping (including light elements) across a wide range of material systems.

Another important direction is the role of neural networks in improving the interpretability and throughput of atomic-resolution tomograms. These approaches not only enhance image quality but also streamline the extraction of structural information, potentially enabling higher-throughput AET workflows. This will make it more feasible to conduct repeated measurements of structural changes under external stimuli, such as thermal annealing, mechanical stress, or electromagnetic fields, thereby facilitating studies of structural dynamics in real time[111,162].

Unsupervised neural networks for denoising low-dose TEM images are also under active development, demonstrating promising results in atomic-resolution tomography under in situ conditions, for example, in liquid-phase TEM where reconstructions are based on the random motion of nanomaterials[110,112,131]. Additionally, deep learning may contribute to uncertainty quantification in atomic models, chemical species classification, and automated interpretation of structure-function relationships.

Despite these promising developments, several challenges remain. Surface atom mobility during acquisition, beam-induced artifacts, and the absence of universally validated ground truth datasets continue to limit the reliability of reconstructed structures[163–165]. Overcoming these obstacles will require ongoing advancements in both algorithm development and experimental methodologies.

Nevertheless, the convergence of deep learning and AET is already reshaping our ability to observe and understand matter at the atomic scale. Continued progress in model interpretability, generalization, and physics-informed training[166–170] will further enhance the reliability, robustness, and accessibility of AET—ultimately making high-precision 3D imaging a routine tool for nanoscience and materials research.

## References

1. Stamenkovic, V. R. *et al.* Improved Oxygen Reduction Activity on Pt3Ni(111) via Increased Surface Site Availability. *Science* **315**, 493–497 (2007).
2. Ding, Y. & Chen, M. Nanoporous Metals for Catalytic and Optical Applications. *MRS Bull.* **34**, 569–576 (2009).
3. Lim, B. *et al.* Pd-Pt Bimetallic Nanodendrites with High Activity for Oxygen Reduction. *Science* **324**, 1302–1305 (2009).
4. Huang, X. *et al.* High-performance transition metal–doped Pt3Ni octahedra for oxygen reduction reaction. *Science* **348**, 1230–1234 (2015).
5. Sharma, N., Ojha, H., Bharadwaj, A., Pathak, D. P. & Sharma, R. K. Preparation and catalytic applications of nanomaterials: a review. *RSC Adv.* **5**, 53381–53403 (2015).
6. Gawande, M. B. *et al.* Cu and Cu-Based Nanoparticles: Synthesis and Applications in Catalysis. *Chem. Rev.* **116**, 3722–3811 (2016).
7. Lin, L. *et al.* Low-temperature hydrogen production from water and methanol using Pt/α-MoC catalysts. *Nature* **544**, 80–83 (2017).
8. Astruc, D. Introduction: Nanoparticles in Catalysis. *Chem. Rev.* **120**, 461–463 (2020).
9. Lee, S. *et al.* A measure of active interfaces in supported catalysts for high-temperature reactions. *Chem* **8**, 815–835 (2022).
10. Law, M., Goldberger, J. & Yang, P. SEMICONDUCTOR NANOWIRES AND NANOTUBES. *Annu. Rev. Mater. Res.* **34**, 83–122 (2004).
11. Tian, B. *et al.* Coaxial silicon nanowires as solar cells and nanoelectronic power sources. *Nature* **449**, 885–889 (2007).
12. Akinwande, D., Petrone, N. & Hone, J. Two-dimensional flexible nanoelectronics. *Nat. Commun.* **5**, 5678 (2014).
13. Dou, L. *et al.* Atomically thin two-dimensional organic-inorganic hybrid perovskites. *Science* **349**, 1518–1521 (2015).
14. Jia, C., Lin, Z., Huang, Y. & Duan, X. Nanowire Electronics: From Nanoscale to Macroscale. *Chem. Rev.* **119**, 9074–9135 (2019).
15. Hossain, N. *et al.* Advances and significances of nanoparticles in semiconductor applications – A review. *Results Eng.* **19**, 101347 (2023).
16. Aricò, A. S., Bruce, P., Scrosati, B., Tarascon, J.-M. & van Schalkwijk, W. Nanostructured materials for advanced energy conversion and storage devices. *Nat. Mater.* **4**, 366–377 (2005).
17. Kang, B. & Ceder, G. Battery materials for ultrafast charging and discharging. *Nature* **458**, 190–193 (2009).
18. Green, M. A., Emery, K., Hishikawa, Y., Warta, W. & Dunlop, E. D. Solar cell efficiency tables (Version 45). *Prog. Photovolt. Res. Appl.* **23**, 1–9 (2015).
19. Pomerantseva, E., Bonaccorso, F., Feng, X., Cui, Y. & Gogotsi, Y. Energy storage: The future enabled by nanomaterials. *Science* **366**, eaan8285 (2019).
20. Sun, H. *et al.* Hierarchical 3D electrodes for electrochemical energy storage. *Nat. Rev. Mater.* **4**, 45–60 (2019).
21. Gohar, O. *et al.* Nanomaterials for advanced energy applications: Recent advancements and future trends. *Mater. Des.* **241**, 112930 (2024).
22. Mohammed, H., Mia, M. F., Wiggins, J. & Desai, S. Nanomaterials for Energy Storage Systems—A Review. *Molecules* **30**, 883 (2025).
23. Qian, X., Liu, J., Fu, L. & Li, J. Quantum spin Hall effect in two-dimensional transition metal dichalcogenides. *Science* **346**, 1344–1347 (2014).
24. Liu, P., Williams, J. R. & Cha, J. J. Topological nanomaterials. *Nat. Rev. Mater.* **4**, 479–496 (2019).
25. Liu, X. & Hersam, M. C. 2D materials for quantum information science. *Nat. Rev. Mater.* **4**, 669–684 (2019).
26. Iyengar, S. A., Puthirath, A. B. & Swaminathan, V. Realizing Quantum Technologies in Nanomaterials and Nanoscience. *Adv. Mater.* **35**, 2107839 (2023).
27. Malik, S. *et al.* Nanomaterials-based biosensor and their applications: A review. *Heliyon* **9**, e19929 (2023).
28. Montblanch, A. R.-P., Barbone, M., Aharonovich, I., Atatüre, M. & Ferrari, A. C. Layered materials as a platform for quantum technologies. *Nat. Nanotechnol.* **18**, 555–571 (2023).
29. Wang, Y., Chen, M., Zhou, F. & Ma, E. High tensile ductility in a nanostructured metal. *Nature* **419**, 912–915 (2002).
30. Lee, J., Mahendra, S. & Alvarez, P. J. J. Nanomaterials in the Construction Industry: A Review of Their Applications and Environmental Health and Safety Considerations. *ACS Nano* **4**, 3580–3590 (2010).
31. Gludovatz, B. *et al.* A fracture-resistant high-entropy alloy for cryogenic applications. *Science* **345**, 1153–1158 (2014).
32. Lilleodden, E. T. & Voorhees, P. W. On the topological, morphological, and microstructural characterization of nanoporous metals. *MRS Bull.* **43**, 20–26 (2018).
33. Mohajerani, A. *et al.* Nanoparticles in Construction Materials and Other Applications, and Implications of Nanoparticle Use. *Materials* **12**, 3052 (2019).
34. Macías-Silva, M. A. *et al.* Nanomaterials in construction industry: An overview of their properties and contributions in building house. *Case Stud. Chem. Environ. Eng.* **10**, 100863 (2024).
35. Rosi, N. L. & Mirkin, C. A. Nanostructures in Biodiagnostics. *Chem. Rev.* **105**, 1547–1562 (2005).
36. Howes, P. D., Chandrawati, R. & Stevens, M. M. Colloidal nanoparticles as advanced biological sensors. *Science* **346**, 1247390 (2014).
37. Huang, X., Zhu, Y. & Kianfar, E. Nano Biosensors: Properties, applications and electrochemical techniques. *J. Mater. Res. Technol.* **12**, 1649–1672 (2021).
38. Altug, H., Oh, S.-H., Maier, S. A. & Homola, J. Advances and applications of nanophotonic biosensors. *Nat. Nanotechnol.* **17**, 5–16 (2022).
39. Ding, Y. *et al.* Gold Nanoparticles for Nucleic Acid Delivery. *Mol. Ther.* **22**, 1075–1083 (2014).
40. Mitchell, M. J. *et al.* Engineering precision nanoparticles for drug delivery. *Nat. Rev. Drug Discov.* **20**, 101–124 (2021).

41. Sultana, A., Zare, M., Thomas, V., Kumar, T. S. S. & Ramakrishna, S. Nano-based drug delivery systems: Conventional drug delivery routes, recent developments and future prospects. *Med. Drug Discov.* **15**, 100134 (2022).
42. Yusuf, A., Almotairy, A. R. Z., Henidi, H., Alshehri, O. Y. & Aldughaim, M. S. Nanoparticles as Drug Delivery Systems: A Review of the Implication of Nanoparticles' Physicochemical Properties on Responses in Biological Systems. *Polymers* **15**, 1596 (2023).
43. Miao, J., Ercius, P. & Billinge, S. J. L. Atomic electron tomography: 3D structures without crystals. *Science* **353**, aaf2157 (2016).
44. Chen, C.-T., Chrzan, D. C. & Gu, G. X. Nano-topology optimization for materials design with atom-by-atom control. *Nat. Commun.* **11**, 3745 (2020).
45. Munkhbat, B. *et al.* Transition metal dichalcogenide metamaterials with atomic precision. *Nat. Commun.* **11**, 4604 (2020).
46. Ji, S., Jun, C., Chen, Y. & Wang, D. Precise Synthesis at the Atomic Scale. *Precis. Chem.* **1**, 199–225 (2023).
47. Haider, M. *et al.* Electron microscopy image enhanced. *Nature* **392**, 768–769 (1998).
48. Krivanek, O. L., Dellby, N. & Lupini, A. R. Towards sub-Å electron beams. *Ultramicroscopy* **78**, 1–11 (1999).
49. Hirsch, P., Cockayne ,D., Spence ,J. & and Whelan, M. 50 Years of TEM of dislocations: Past, present and future. *Philos. Mag.* **86**, 4519–4528 (2006).
50. Smith, D. J. Development of Aberration-Corrected Electron Microscopy. *Microsc. Microanal.* **14**, 2–15 (2008).
51. Rose, H. H. Historical aspects of aberration correction. *J. Electron Microsc. (Tokyo)* **58**, 77–85 (2009).
52. Erni, R., Rossell, M. D., Kisielowski, C. & Dahmen, U. Atomic-Resolution Imaging with a Sub-50-pm Electron Probe. *Phys. Rev. Lett.* **102**, 096101 (2009).
53. Saghi, Z. & Midgley, P. A. Electron Tomography in the (S)TEM: From Nanoscale Morphological Analysis to 3D Atomic Imaging. *Annu. Rev. Mater. Res.* **42**, 59–79 (2012).
54. Hosokawa, F., Sawada, H., Kondo, Y., Takayanagi, K. & Suenaga, K. Development of Cs and Cc correctors for transmission electron microscopy. *Microscopy* **62**, 23–41 (2013).
55. Midgley, P. A. & Dunin-Borkowski, R. E. Electron tomography and holography in materials science. *Nat. Mater.* **8**, 271–280 (2009).
56. Williams, D. B. & Carter, C. B. *Transmission Electron Microscopy*. (Springer US, Boston, MA, 2009). doi:10.1007/978-0-387-76501-3.
57. Zečević, J., de Jong, K. P. & de Jongh, P. E. Progress in electron tomography to assess the 3D nanostructure of catalysts. *Curr. Opin. Solid State Mater. Sci.* **17**, 115–125 (2013).
58. Ercius, P., Alaidi, O., Rames, M. J. & Ren, G. Electron Tomography: A Three-Dimensional Analytic Tool for Hard and Soft Materials Research. *Adv. Mater.* **27**, 5638–5663 (2015).
59. Neumüller, J. Electron tomography—a tool for ultrastructural 3D visualization in cell biology and histology. *Wien. Med. Wochenschr.* **168**, 322–329 (2018).
60. Lee, J., Jeong, C., Lee, T., Ryu, S. & Yang, Y. Direct Observation of Three-Dimensional Atomic Structure of Twinned Metallic Nanoparticles and Their Catalytic Properties. *Nano Lett.* **22**, 665–672 (2022).
61. Tersoff, J. & Hamann, D. R. Theory of the scanning tunneling microscope. *Phys. Rev. B* **31**, 805–813 (1985).
62. Hofer, W. A., Foster, A. S. & Shluger, A. L. Theories of scanning probe microscopes at the atomic scale. *Rev. Mod. Phys.* **75**, 1287–1331 (2003).
63. Giessibl, F. J. Advances in atomic force microscopy. *Rev. Mod. Phys.* **75**, 949–983 (2003).
64. Jalili, N. & Laxminarayana, K. A review of atomic force microscopy imaging systems: application to molecular metrology and biological sciences. *Mechatronics* **14**, 907–945 (2004).
65. Su, S., Zhao, J. & Ly, T. H. Scanning Probe Microscopies for Characterizations of 2D Materials. *Small Methods* **8**, 2400211 (2024).
66. Barberán, J. R. M. Scanning Tunneling Microscopy: A Review. *Nanotechnol. Percept.* 1994–2006 (2024) doi:10.62441/nano-ntp.vi.3063.
67. Ladd, M. & Palmer, R. *Structure Determination by X-Ray Crystallography*. (Springer US, Boston, MA, 2003). doi:10.1007/978-1-4615-0101-5.
68. *Principles of Protein X-Ray Crystallography*. (Springer, New York, NY, 2007). doi:10.1007/0-387-33746-6.
69. Kelly, A. & Knowles, K. M. *Crystallography and Crystal Defects*. (Wiley, MA, Chichester, West Sussex, UK ; Malden, 2012).
70. McNally, P. J. 3D imaging of crystal defects. *Nature* **496**, 37–38 (2013).
71. Gemmi, M. *et al.* 3D Electron Diffraction: The Nanocrystallography Revolution. *ACS Cent. Sci.* **5**, 1315–1329 (2019).
72. Losko, A. S. & Vogel, S. C. 3D isotope density measurements by energy-resolved neutron imaging. *Sci. Rep.* **12**, 6648 (2022).
73. Kutsal, M., Poulsen, H. F., Winther, G., Sørensen, H. O. & Detlefs, C. High-resolution 3D X-ray diffraction microscopy: 3D mapping of deformed metal microstructures. *J. Appl. Crystallogr.* **55**, 1125–1138 (2022).
74. Hull, D. & Bacon, D. J. *Introduction to Dislocations*. (Elsevier, 2011).
75. Miao, J., Charalambous, P., Kirz, J. & Sayre, D. Extending the methodology of X-ray crystallography to allow imaging of micrometre-sized non-crystalline specimens. *Nature* **400**, 342–344 (1999).
76. Aert, S. V., Dyck, D. V. & Dekker, A. J. den. Resolution of coherent and incoherent imaging systems reconsidered - Classical criteria and a statistical alternative. *Opt. Express* **14**, 3830–3839 (2006).
77. Huang, W. J. *et al.* Coordination-dependent surface atomic contraction in nanocrystals revealed by coherent diffraction. *Nat. Mater.* **7**, 308–313 (2008).
78. Robinson, I. & Harder, R. Coherent X-ray diffraction imaging of strain at the nanoscale. *Nat. Mater.* **8**, 291–298 (2009).

79. Miao, J., Ishikawa, T., Robinson, I. K. & Murnane, M. M. Beyond crystallography: Diffractive imaging using coherent x-ray light sources. *Science* **348**, 530–535 (2015).
80. Aidukas, T. *et al.* High-performance 4-nm-resolution X-ray tomography using burst ptychography. *Nature* **632**, 81–88 (2024).
81. Miao, J. Computational microscopy with coherent diffractive imaging and ptychography. *Nature* **637**, 281–295 (2025).
82. Kelly, T. F. & Miller, M. K. Atom probe tomography. *Rev. Sci. Instrum.* **78**, 031101 (2007).
83. Miller, M. K., Kelly, Thomas. F., Rajan, K. & Ringer, S. P. The future of atom probe tomography. *Mater. Today* **15**, 158–165 (2012).
84. Moody, M. P. *et al.* Atomically resolved tomography to directly inform simulations for structure–property relationships. *Nat. Commun.* **5**, 5501 (2014).
85. Peng, Z. *et al.* On the detection of multiple events in atom probe tomography. *Ultramicroscopy* **189**, 54–60 (2018).
86. Gault, B. *et al.* Atom probe tomography. *Nat. Rev. Methods Primer* **1**, 1–30 (2021).
87. De Rosier, D. J. & Klug, A. Reconstruction of Three Dimensional Structures from Electron Micrographs. *Nature* **217**, 130–134 (1968).
88. Hoppe, W., Langer, R., Knesch, G. & Poppe, Ch. Protein-Kristallstrukturanalyse mit Elektronenstrahlen. *Naturwissenschaften* **55**, 333–336 (1968).
89. Hart, R. G. Electron Microscopy of Unstained Biological Material: The Polytropic Montage. *Science* **159**, 1464–1467 (1968).
90. Scott, M. C. *et al.* Electron tomography at 2.4-ångström resolution. *Nature* **483**, 444–447 (2012).
91. Goris, B. *et al.* Measuring Lattice Strain in Three Dimensions through Electron Microscopy. *Nano Lett.* **15**, 6996–7001 (2015).
92. Haberfehlner, G. *et al.* Formation of bimetallic clusters in superfluid helium nanodroplets analysed by atomic resolution electron tomography. *Nat. Commun.* **6**, 8779 (2015).
93. Xu, R. *et al.* Three-dimensional coordinates of individual atoms in materials revealed by electron tomography. *Nat. Mater.* **14**, 1099–1103 (2015).
94. Miao, J., Förster, F. & Levi, O. Equally sloped tomography with oversampling reconstruction. *Phys. Rev. B* **72**, 052103 (2005).
95. Yang, Y. *et al.* Deciphering chemical order/disorder and material properties at the single-atom level. *Nature* **542**, 75–79 (2017).
96. Wang, C. *et al.* Three-Dimensional Atomic Structure of Grain Boundaries Resolved by Atomic-Resolution Electron Tomography. *Matter* **3**, 1999–2011 (2020).
97. Chen, C.-C. *et al.* Three-dimensional imaging of dislocations in a nanoparticle at atomic resolution. *Nature* **496**, 74–77 (2013).
98. Sun, Z. *et al.* Strain release by 3D atomic misfit in fivefold twinned icosahedral nanoparticles with amorphization and dislocations. *Nat. Commun.* **16**, 1595 (2025).
99. Tian, X. *et al.* Correlating the three-dimensional atomic defects and electronic properties of two-dimensional transition metal dichalcogenides. *Nat. Mater.* **19**, 867–873 (2020).
100. Tian, X. *et al.* Capturing 3D atomic defects and phonon localization at the 2D heterostructure interface. *Sci. Adv.* **7**, eabi6699.
101. Jeong, C. *et al.* Revealing the three-dimensional arrangement of polar topology in nanoparticles. *Nat. Commun.* **15**, 3887 (2024).
102. Hong, J. *et al.* Metastable hexagonal close-packed palladium hydride in liquid cell TEM. *Nature* **603**, 631–636 (2022).
103. Jo, H. *et al.* Direct strain correlations at the single-atom level in three-dimensional core-shell interface structures. *Nat. Commun.* **13**, 5957 (2022).
104. Li, Z. *et al.* Probing the atomically diffuse interfaces in Pd@Pt core-shell nanoparticles in three dimensions. *Nat. Commun.* **14**, 2934 (2023).
105. Zhang, Y. *et al.* Three-dimensional atomic insights into the metal-oxide interface in Zr-ZrO2 nanoparticles. *Nat. Commun.* **15**, 7624 (2024).
106. Yang, Y. *et al.* Determining the three-dimensional atomic structure of an amorphous solid. *Nature* **592**, 60–64 (2021).
107. Yuan, Y. *et al.* Three-dimensional atomic packing in amorphous solids with liquid-like structure. *Nat. Mater.* **21**, 95–102 (2022).
108. Lee, J., Jeong, C. & Yang, Y. Single-atom level determination of 3-dimensional surface atomic structure via neural network-assisted atomic electron tomography. *Nat. Commun.* **12**, 1962 (2021).
109. Yang, Y. *et al.* Atomic-scale identification of active sites of oxygen reduction nanocatalysts. *Nat. Catal.* **7**, 796–806 (2024).
110. Kim, B. H. *et al.* Critical differences in 3D atomic structure of individual ligand-protected nanocrystals in solution. *Science* **368**, 60–67 (2020).
111. Zhou, J. *et al.* Observing crystal nucleation in four dimensions using atomic electron tomography. *Nature* **570**, 500–503 (2019).
112. Kang, S. *et al.* Time-resolved Brownian tomography of single nanocrystals in liquid during oxidative etching. *Nat. Commun.* **16**, 1158 (2025).
113. Zhou, J., Yang, Y., Ercius, P. & Miao, J. Atomic electron tomography in three and four dimensions. *MRS Bull.* **45**, 290–297 (2020).
114. Van Aert, S., Batenburg, K. J., Rossell, M. D., Erni, R. & Van Tendeloo, G. Three-dimensional atomic imaging of crystalline nanoparticles. *Nature* **470**, 374–377 (2011).
115. Goris, B. *et al.* Three-Dimensional Elemental Mapping at the Atomic Scale in Bimetallic Nanocrystals. *Nano Lett.* **13**, 4236–4241 (2013).

116. Dai, Y. *et al.* Mapping Surface and Subsurface Atomic Structures of Au@Pd Core–Shell Nanoparticles in Three Dimensions. *ACS Nano* **19**, 9006–9016 (2025).
117. Leary, R. K. & Midgley, P. A. Electron Tomography in Materials Science. in *Springer Handbook of Microscopy* (eds. Hawkes, P. W. & Spence, J. C. H.) 1279–1329 (Springer International Publishing, Cham, 2019). doi:10.1007/978-3-030-00069-1_26.
118. Kawase, N., Kato, M., Nishioka, H. & Jinnai, H. Transmission electron microtomography without the "missing wedge" for quantitative structural analysis. *Ultramicroscopy* **107**, 8–15 (2007).
119. Friedrich, H., de Jongh, P. E., Verkleij, A. J. & de Jong, K. P. Electron Tomography for Heterogeneous Catalysts and Related Nanostructured Materials. *Chem. Rev.* **109**, 1613–1629 (2009).
120. Ni, B. & Wang, X. Face the Edges: Catalytic Active Sites of Nanomaterials. *Adv. Sci.* **2**, 1500085 (2015).
121. Lin, Y. *et al.* Adhesion and Atomic Structures of Gold on Ceria Nanostructures: The Role of Surface Structure and Oxidation State of Ceria Supports. *Nano Lett.* **15**, 5375–5381 (2015).
122. Liu, F., Wu, C., Ding, X. & Sun, J. Atomic modification of Mo(1 0 0) surface for corrosion resistance. *Appl. Surf. Sci.* **610**, 155509 (2023).
123. Hasegawa, S., Tong, X., Takeda, S., Sato, N. & Nagao, T. Structures and electronic transport on silicon surfaces. *Prog. Surf. Sci.* **60**, 89–257 (1999).
124. Kolmer, M. *et al.* Electronic transport in planar atomic-scale structures measured by two-probe scanning tunneling spectroscopy. *Nat. Commun.* **10**, 1573 (2019).
125. Wang, Z. L. Transmission Electron Microscopy of Shape-Controlled Nanocrystals and Their Assemblies. *J. Phys. Chem. B* **104**, 1153–1175 (2000).
126. Tian, Y. *et al.* Fast coalescence of metallic glass nanoparticles. *Nat. Commun.* **10**, 5249 (2019).
127. Yang, W. *et al.* Deep Learning for Single Image Super-Resolution: A Brief Review. *IEEE Trans. Multimed.* **21**, 3106–3121 (2019).
128. Yang, S.-H. *et al.* Deep Learning-Assisted Quantification of Atomic Dopants and Defects in 2D Materials. *Adv. Sci.* **8**, 2101099 (2021).
129. Cho, I. *et al.* Deep-learning-based gas identification by time-variant illumination of a single micro-LED-embedded gas sensor. *Light Sci. Appl.* **12**, 95 (2023).
130. Yao, L., Lyu, Z., Li, J. & Chen, Q. No ground truth needed: unsupervised sinogram inpainting for nanoparticle electron tomography (UsiNet) to correct missing wedges. *Npj Comput. Mater.* **10**, 1–14 (2024).
131. Kim, J. *et al.* Self-supervised machine learning framework for high-throughput electron microscopy. *Sci. Adv.* **11**, eads5552 (2025).
132. Ding, G., Liu, Y., Zhang, R. & Xin, H. L. A joint deep learning model to recover information and reduce artifacts in missing-wedge sinograms for electron tomography and beyond. *Sci. Rep.* **9**, 12803 (2019).
133. Wang, C., Ding, G., Liu, Y. & Xin, H. L. 0.7 Å Resolution Electron Tomography Enabled by Deep-Learning-Aided Information Recovery. *Adv. Intell. Syst.* **2**, 2000152 (2020).
134. Yu, Y. *et al.* Ensemble Cross UNet Transformers for Augmentation of Atomic Electron Tomography. *IEEE Trans. Instrum. Meas.* **73**, 1–14 (2024).
135. Iwai, H. *et al.* Atomic-Scale 3D Structure of a Supported Pd Nanoparticle Revealed by Electron Tomography with Convolution Neural Network-Based Image Inpainting. *Small Methods* **8**, 2301163 (2024).
136. Radermacher, M. Weighted Back-projection Methods. in *Electron Tomography: Methods for Three-Dimensional Visualization of Structures in the Cell* (ed. Frank, J.) 245–273 (Springer, New York, NY, 2006). doi:10.1007/978-0-387-69008-7_9.
137. Andersen, A. H. & Kak, A. C. Simultaneous Algebraic Reconstruction Technique (SART): A superior implementation of the ART algorithm. *Ultrason. Imaging* **6**, 81–94 (1984).
138. Çiçek, Ö., Abdulkadir, A., Lienkamp, S. S., Brox, T. & Ronneberger, O. 3D U-Net: Learning Dense Volumetric Segmentation from Sparse Annotation. in *Medical Image Computing and Computer-Assisted Intervention – MICCAI 2016* (eds. Ourselin, S., Joskowicz, L., Sabuncu, M. R., Unal, G. & Wells, W.) vol. 9901 424–432 (Springer International Publishing, Cham, 2016).
139. Pryor, A. *et al.* GENFIRE: A generalized Fourier iterative reconstruction algorithm for high-resolution 3D imaging. *Sci. Rep.* **7**, 10409 (2017).
140. Schneider, S., Surrey, A., Pohl, D., Schultz, L. & Rellinghaus, B. Atomic surface diffusion on Pt nanoparticles quantified by high-resolution transmission electron microscopy. *Micron* **63**, 52–56 (2014).
141. Lee, J.-W. *et al.* Conducting interfaces between LaAlO3 and thick homoepitaxial SrTiO3 films for transferable templates. *Appl. Surf. Sci.* **582**, 152480 (2022).
142. Kim, H.-S., An, J.-S., Bae, H. B. & Chung, S.-Y. Atomic-scale observation of premelting at 2D lattice defects inside oxide crystals. *Nat. Commun.* **14**, 2255 (2023).
143. Yu, I. C. *et al.* Study on microscopic origin of spin excitation in TmFeO3 by time-resolved multicolor optical pump-probe spectroscopy. *Curr. Appl. Phys.* **62**, 54–59 (2024).
144. Ophus, C., Ercius, P., Sarahan, M., Czarnik, C. & Ciston, J. Recording and Using 4D-STEM Datasets in Materials Science. *Microsc. Microanal.* **20**, 62–63 (2014).
145. Ophus, C. Four-Dimensional Scanning Transmission Electron Microscopy (4D-STEM): From Scanning Nanodiffraction to Ptychography and Beyond. *Microsc. Microanal.* **25**, 563–582 (2019).
146. Jiang, Y. *et al.* Electron ptychography of 2D materials to deep sub-ångström resolution. *Nature* **559**, 343–349 (2018).
147. Chen, Z. *et al.* Mixed-state electron ptychography enables sub-angstrom resolution imaging with picometer precision at low dose. *Nat. Commun.* **11**, 2994 (2020).
148. Chen, Z. *et al.* Electron ptychography achieves atomic-resolution limits set by lattice vibrations. *Science* **372**, 826–831 (2021).

149. Yang, W., Sha, H., Cui, J., Mao, L. & Yu, R. Local-orbital ptychography for ultrahigh-resolution imaging. *Nat. Nanotechnol.* **19**, 612–617 (2024).
150. Dong, Z. *et al.* Visualization of oxygen vacancies and self-doped ligand holes in La3Ni2O7−δ. *Nature* **630**, 847–852 (2024).
151. Ribet, S. M. *et al.* Uncovering the three-dimensional structure of upconverting core–shell nanoparticles with multislice electron ptychography. *Appl. Phys. Lett.* **124**, 240601 (2024).
152. Dong, Z. *et al.* Sub-nanometer depth resolution and single dopant visualization achieved by tilt-coupled multislice electron ptychography. *Nat. Commun.* **16**, 1219 (2025).
153. Van den Broek, W. & Koch, C. T. Method for Retrieval of the Three-Dimensional Object Potential by Inversion of Dynamical Electron Scattering. *Phys. Rev. Lett.* **109**, 245502 (2012).
154. Van den Broek, W. & Koch, C. T. General framework for quantitative three-dimensional reconstruction from arbitrary detection geometries in TEM. *Phys. Rev. B* **87**, 184108 (2013).
155. Chang, D. J. *et al.* Ptychographic atomic electron tomography: Towards three-dimensional imaging of individual light atoms in materials. *Phys. Rev. B* **102**, 174101 (2020).
156. Lee, J., Lee, M., Park, Y., Ophus, C. & Yang, Y. Multislice Electron Tomography Using Four-Dimensional Scanning Transmission Electron Microscopy. *Phys. Rev. Appl.* **19**, 054062 (2023).
157. Pelz, P. M. *et al.* Solving complex nanostructures with ptychographic atomic electron tomography. *Nat. Commun.* **14**, 7906 (2023).
158. Romanov, A., Cho, M. G., Scott, M. C. & Pelz, P. Multi-slice electron ptychographic tomography for three-dimensional phase-contrast microscopy beyond the depth of focus limits. *J. Phys. Mater.* **8**, 015005 (2024).
159. You, S., Romanov, A. & Pelz, P. M. Near-isotropic sub-Ångstrom 3d resolution phase contrast imaging achieved by end-to-end ptychographic electron tomography. *Phys. Scr.* **100**, 015404 (2024).
160. Ren, D., Ophus, C., Chen, M. & Waller, L. A multiple scattering algorithm for three dimensional phase contrast atomic electron tomography. *Ultramicroscopy* **208**, 112860 (2020).
161. Lee, J. *et al.* PhaseT3M: 3D Imaging at 1.6 Å Resolution via Electron Cryo-Tomography with Nonlinear Phase Retrieval. Preprint at https://doi.org/10.48550/arXiv.2504.16332 (2025).
162. Jeong, C. *et al.* Atomic-scale 3D structural dynamics and functional degradation of Pt alloy nanocatalysts. Preprint at https://doi.org/10.48550/arXiv.2411.01727 (2024).
163. Isaacson, M. Electron beam induced damage of organic solids: Implications for analytical electron microscopy. *Ultramicroscopy* **4**, 193–199 (1979).
164. Zhang, D. *et al.* Atomic-resolution transmission electron microscopy of electron beam–sensitive crystalline materials. *Science* **359**, 675–679 (2018).
165. Xu, X. *et al.* Unravelling nonclassical beam damage mechanisms in metal-organic frameworks by low-dose electron microscopy. *Nat. Commun.* **16**, 261 (2025).
166. Karniadakis, G. E. *et al.* Physics-informed machine learning. *Nat. Rev. Phys.* **3**, 422–440 (2021).
167. Ha, S. & Jeong, H. Unraveling hidden interactions in complex systems with deep learning. *Sci. Rep.* **11**, 12804 (2021).
168. Wu, Y., Sicard, B. & Gadsden, S. A. Physics-informed machine learning: A comprehensive review on applications in anomaly detection and condition monitoring. *Expert Syst. Appl.* **255**, 124678 (2024).
169. Vasiliauskaite, V. & Antulov-Fantulin, N. Generalization of neural network models for complex network dynamics. *Commun. Phys.* **7**, 1–10 (2024).
170. Pateras, J., Zhang, C., Majumdar, S., Pal, A. & Ghosh, P. Physics-informed machine learning for automatic model reduction in chemical reaction networks. *Sci. Rep.* **15**, 7980 (2025).